\documentclass{Interspeech2024}
\usepackage{cite}
\usepackage{url}
\usepackage{makecell, hhline}
\usepackage{float}% nidanfloat}
\usepackage{amsmath,amssymb,amsfonts}
\usepackage{algorithmic}
\usepackage{graphicx}
\usepackage{textcomp}
\usepackage{xcolor}
\usepackage{authblk}
\usepackage[nolist,nohyperlinks]{acronym}
\usepackage{multirow}
\usepackage[subscriptcorrection,varg,varvw,cmbraces,cmintegrals]{newtxmath}

\usepackage[style=base]{caption}
\interspeechcameraready

\title{XANE: eXplainable Acoustic Neural Embeddings}

\name[affiliation={1}]{Sri Harsha}{Dumpala}
\name[affiliation={2}]{Dushyant}
{Sharma}
\name[affiliation={1}]{Chandramouli Shama}{Sastri}
\name[affiliation={2}]{Stanislav}{Kruchinin}

\name[affiliation={2}]{James}{Fosburgh}
\name[affiliation={3}]{Patrick A.}{Naylor}

\address{
  $^1$Vector Institute, Canada\\
  $^2$Microsoft Inc. \\
  $^3$Imperial College, UK}
\email{dusharma@microsoft.com}


\begin{document}

\maketitle

\begin{abstract}
We present a novel method for extracting neural embeddings that model the background acoustics of a speech signal. The extracted embeddings are used to estimate specific parameters related to the background acoustic properties of the signal in a non-intrusive manner, which allows the embeddings to be explainable in terms of those parameters. We illustrate the value of these embeddings by performing clustering experiments on unseen test data and show that the proposed embeddings achieve a mean F1 score of 95.2\% for three different tasks, outperforming significantly the WavLM based signal embeddings. We also show that the proposed method can explain the embeddings by estimating 14 acoustic parameters characterizing the background acoustics, including reverberation and noise levels, overlapped speech detection, CODEC type detection and noise type detection with high accuracy and a real-time factor 17 times lower than an external baseline method.
\end{abstract}

\section{Introduction}\label{sec:intro}
Non-intrusive assessment of the background acoustics of a speech signal enables the development and deployment of more robust speech processing systems such as Automatic Speech Recognition~(ASR). In particular, the use case of distant ASR in an enclosed space (e.g. a room or car) is of great interest as it allows for a more natural, hands-free interaction between humans and machines. In such use cases however, high levels of reverberation and noise may be introduced into the recorded signal. Reverberation has been shown to significantly degrade the performance of ASR systems that take no measures to compensate for it~\cite{haeb2020far}. Modern ASR systems are based on deep neural networks~(DNNs) and one way to make them more robust is through data augmentation~\cite{nisa_eusipco23,7953152}. A speech signal may also be adversely affected by additive environmental noise and CODEC artifacts, and previous studies have shown that accurate estimation of these parameters can be beneficial for ASR~\cite{nisapp,parada2015reverberant}, de-reverberation~\cite{williamson2017time}, text-to-speech~(TTS)~\cite{kitawaki1988quality} and speaker diarization~\cite{weychan2014improving,hu2015speaker}. The effect of acoustic reverberation can be modeled as the convolution between a clean~(anechoic) speech signal and a Room Impulse Response~(RIR)~\cite{naylor2010speech}. A number of techniques for simulating an RIR have been proposed and most require the room volume and reflection coefficient parameters to be defined~\cite{nisa_eusipco23}. Although the RIR captures all the information about the reverberation, it is in some cases beneficial to estimate a small number of parameters that characterize the RIR. These include the Clarity index~($C_{50}$), reverberation time~($T_{60}$) and the direct-to-reverberant ratio~(DRR)~\cite{parada2014non}. In~\cite{hu2015speaker}, it was shown that the $C_{5}$ metric is valuable for speaker diarization. The noise level is typically characterized by the Signal-To-Noise Ratio~(SNR) and the combination of all degrading effects can measured from a perceptual quality and intelligibly perspective by methods such as PESQ~\cite{recommendation2001perceptual} and ESTOI~\cite{estoi}. The problem of Voice Activity Detection~(VAD) is also closely linked with the accurate estimation of background acoustic parameters. Over the past decade, a number of algorithms have been proposed for the task of non-intrusive signal analysis~\cite{parada2014non, Sharma2020-NIE, Dsharma2016, Gamper2018}, including methods for estimating reverberation parameters~\cite{parada2014non, Gamper2018, Sharma2020-NIE}, objective speech quality and intelligibility~\cite{sharma2019non, mittag2021nisqa, yi2022conferencingspeech} and the bit rate of speech CODECs~\cite{sharma2017non}. The non-intrusive prediction of perceptual speech quality in terms of the Mean~Opinion~Score~(MOS) has attained a lot of interest in recent years~\cite{yi2022conferencingspeech, gamper2019intrusive}. More recently, a number of methods have been proposed that estimate some room acoustic characteristics along with MOS~\cite{hajal2022mosra, kumar2023torchaudio, nisa_eusipco23, el2023efficient}. In~\cite{kumar2023torchaudio}, the authors propose a Transformer-based model to estimate PESQ, STOI and Scale-Invariant Signal-to-Distortion Ratio~(SI-SDR). In~\cite{el2023efficient}, a Transformer-based model was used to estimate SNR, speech transmission index~(STI), $T_{60}$, DRR and $C_{50}$. Hajal~et.~al.~\cite{hajal2022mosra} propose the BYOL-S with Convolutional Vision Transformer~(CvT) model to estimate these metrics. In this work, we propose XANE, which extracts a neural embedding that encapsulates information about the background acoustics in a speech signal in the form of a vector representation. XANE operates in a non-intrusive framework and makes the embeddings explainable by further estimating a wide range of background acoustic parameters from the neural embeddings. We experiment with Transformer and Conformer based neural architectures using the well-established Mel Filterbank features and a recent self-supervised learning based representation known as WavLM~\cite{chen2022wavlm}. We also implemented the NISA+~\cite{nisapp} algorithm in PyTorch and trained it on our data as a state of the art baseline method. Our key contributions are (1) a novel neural embedding that encapsulates properties of the background acoustics that is made explainable by a (2) multi-task parameter estimator that models 14 parameters including room reverberation and noise. We show that (3) the use of self-supervised learning based WavLM features do not add value over the conventional Mel filter bank~(MelFB) features and finally, (4) we show that the Transformer and Conformer models outperform the baseline models across all the tasks, in terms of accuracy and run time performance. The remainder of the paper is organized as follows. We present the XANE system in Section~\ref{sec:nisa-emb} with experiments and data in Section~\ref{sec:experiments} and \ref{sec:data}, followed by results in Section~\ref{sec:res} and conclusions in Section~\ref{sec:conclusions}.
\section{XANE System}\label{sec:nisa-emb}
\begin{figure}[H]%[!tbp]%[htbp]
\centering
  \centerline{\includegraphics[width=0.99\linewidth]{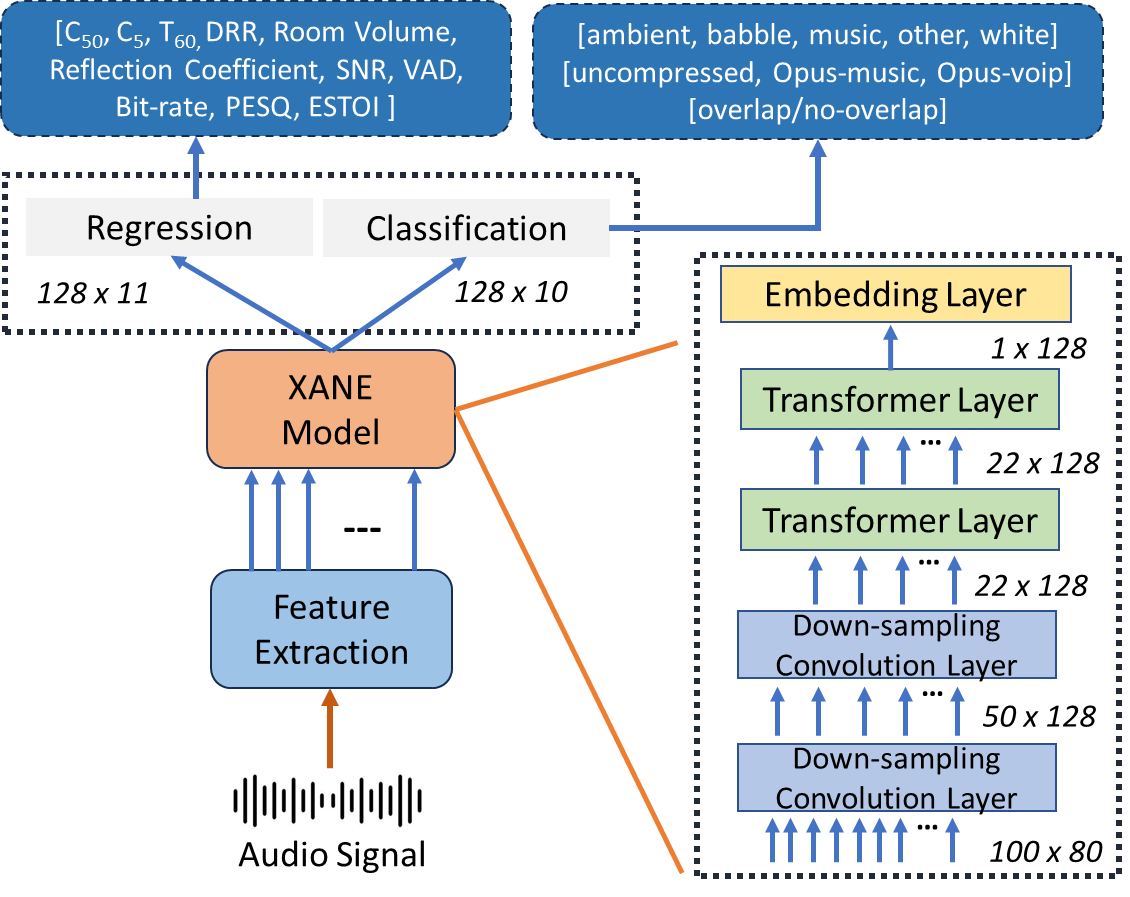}}
  \caption{XANE architecture based on Transformers and MelFB features.}
  \label{fig:nisa_outline}
  \vspace*{-15pt}
\end{figure}

Figure \ref{fig:nisa_outline} shows the architecture of XANE. In the following we describe the Feature extraction and Model architecture components in more detail.
\subsection{Feature extraction} 
\vspace*{-7pt}
We experimented with two types of features: 1) traditional MelFB coefficients and 2) WavLM coefficients based on self-supervised learning. Both MelFB and WavLM features are mean and variance normalized before further processing. All signals were sampled at 16 kHz and 80 dimension MelFB features were extracted using a frame size of 25~ms and 10~ms frame increment. The WavLM features are neural embeddings that we extracted from a pre-trained model~\cite{chen2022wavlm} that was trained on 94k hours of English audio, featuring diverse speakers and scenarios, that takes raw speech as input and outputs an embedding of 768 dimension with a 20~ms frame increment (the default for WavLM). 
\vspace*{-5pt}
\subsection{Model Architectures}
\vspace*{-7pt}
We experimented with the state of the art, transformer and conformer model architectures for XANE. In each case, a segment of 1.0~s duration was used (corresponding to 100 frames for the MelFB and 50 for WavLM features). Also common to all model architectures is the format of the output layer, consisting of 11 linear units (one for each regression task), and three separate softmax blocks for the 3 different classification tasks. The softmax block corresponding to noise type classification (comprising ambient, babble, music, other and white noise), CODEC type classification (uncompressed, Opus ``music'' and ``speech' presets) and speech overlap detection (overlapped or non-overlapped) consists of 5, 3 and 2 units, respectively. The regression tasks include 6 tasks related to room reverberation in addition to SNR, VAD, PESQ~\cite{recommendation2001perceptual}, ESTOI~\cite{estoi} and CODEC bit rate. The reverberation metrics include $C_{50}$, $C_{5}$, $T_{60}$, DRR, room volume and reflection coefficient. The ground truth for the reverberation metrics was computed from RIRs simulated using the image method~\cite{Allen1979} and the ground truth SNR was computed on a per-segment basis (1 second length). The ground truth VAD labels were obtained by an energy based VAD from the clean speech utterances, in 10~ms frames and averaged over the 1~s segments. The ground truth PESQ and ESTOI scores were obtained by applying clean and degraded audio to the respective intrusive systems.
\begin{figure}[H] %[!tbp]%[htbp]
\centering
  \includegraphics[width=0.9\linewidth]{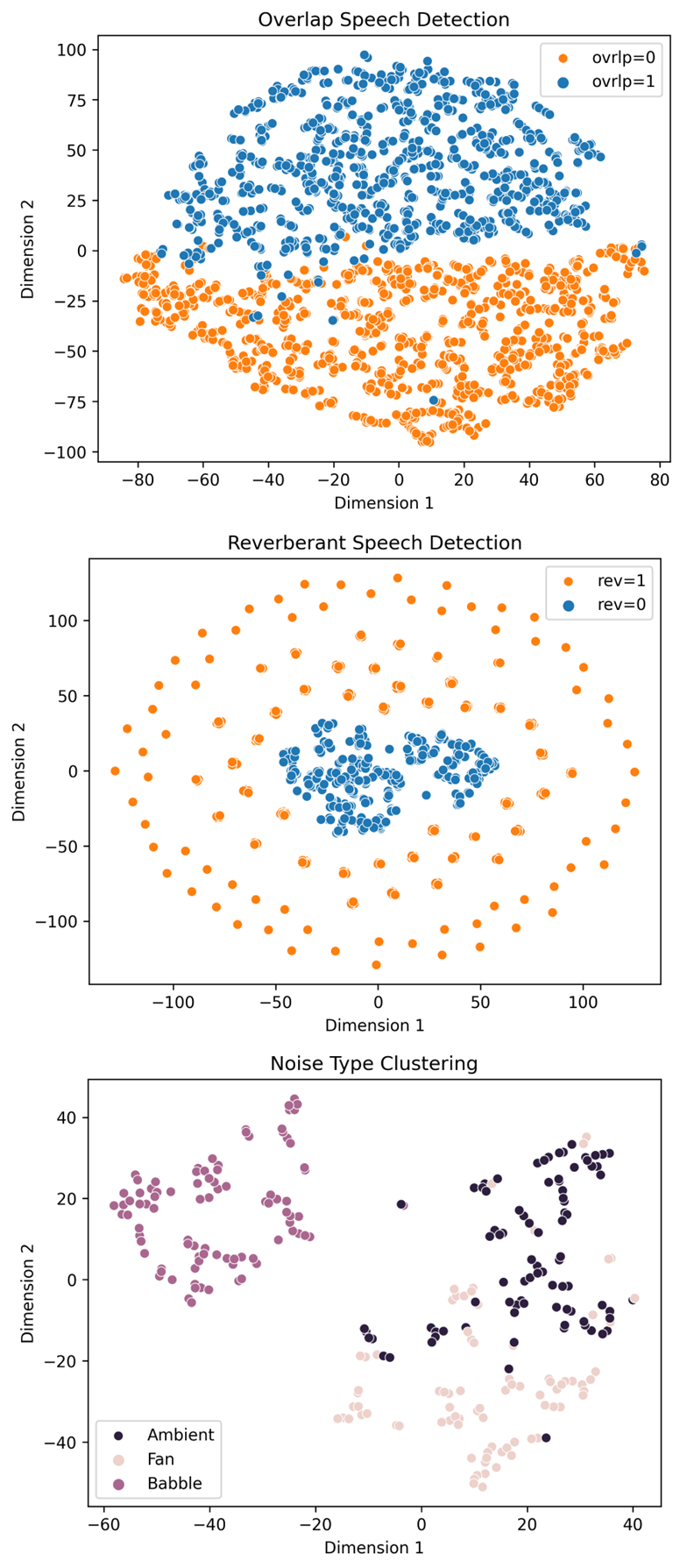}
  \caption{T-SNE~\cite{tsne} plots for XANE embeddings (top: overlap detection, middle: reverberant speech detection and bottom: noisy type clustering) on the ACE data.}
  \label{fig:tsne_conformer}
  \vspace*{-7pt}
\end{figure}
\hspace*{-15pt}In the XANE architecture, the penultimate layer is designated as the embedding layer, based on which acoustic parameters are estimated as outlined in Fig.~\ref{fig:nisa_outline}. The embedding is a vector of dimension 128 and is made explainable by extracting the specific acoustic parameters. The Transformer model consists of a downsampling convolutional block followed by a Transformer block. The downsampling block consists of two convolutional~(Conv.) layers which reduce the input frame rate by a factor of 4. The output of the downsampling blocks is processed through a Transformer block with two encoder layers. For MelFB, each of the convolutional layers have 256 channels with a stride of 2. We use a Transformer layer with 256-dimensional input, 8 attention heads and a 256-dimensional fully connected linear layer. For WavLM, we use Conv. layers each with 128 channels and a stride of (2, 1). We use a Transformer layer with 128-d input, 8 attention heads and a 256-d fully connected, linear layer. The output of the Transformer block is passed through a fully connected layer with 128 units (the embedding layer) and the GELU activation function~\cite{Hendrycks2016Jun}, followed by the output layer. The architecture of the Conformer model is same as the Transformer model, except the Transformer layers were replaced by Conformer layers. The Conformer model consists of a down-sampling convolutional block followed by a Transformer block. The downsampling block consists of two convolutional layers which downsample the input frame rate by a factor of 4. The output of the downsampling block is passed through a conformer block with two conformer layers. The output of the Conformer block is passed through a fully connected layer with 128 units (embedding layer) and the GELU activation function, followed by the output layer. 
\begin{table*}
\caption{Parameter estimation (explaining the embeddings) on VCTK test set (* 92 MAE for TAS is for STOI labels).}
\label{vctk_results}
\centering
\begin{tabular}{|c|l|c|c|c|c|c|c|c|c|c|}
\cline{3-11} \cline{4-11} \cline{5-11} \cline{6-11} \cline{7-11} \cline{8-11} \cline{9-11} \cline{10-11} \cline{11-11} 
\multicolumn{1}{c}{} &  & \multicolumn{2}{c|}{\textbf{XANE-MelFB}} & \multicolumn{2}{c|}{\textbf{XANE-MelFB-NN}} & \multicolumn{2}{c|}{\textbf{XANE-WavLM}} & \multirow{2}{*}{\textbf{NISA}} & \multirow{2}{*}{\textbf{TAS}} & \multirow{2}{*}{\textbf{SN}}\tabularnewline
\cline{3-8} \cline{4-8} \cline{5-8} \cline{6-8} \cline{7-8} \cline{8-8} 
\multicolumn{1}{c}{} &  & \textbf{Trans.} & \textbf{Conf.} & \textbf{Trans.} & \textbf{Conf.} & \textbf{Trans.} & \textbf{Conf.} &  &  & \tabularnewline
\hline 
\multirow{10}{*}{\textbf{MAE}} & \textbf{C50 (dB)} & 3.2 & 2.9 & 2.9 & \textbf{2.7} & 3.3 & 3.3 & 3.1 & - & -\tabularnewline
\cline{2-11} \cline{3-11} \cline{4-11} \cline{5-11} \cline{6-11} \cline{7-11} \cline{8-11} \cline{9-11} \cline{10-11} \cline{11-11} 
 & \textbf{T60 (ms)} & 112 & \textbf{82} & 126 & 88 & 97 & 97 & 86 & - & -\tabularnewline
\cline{2-11} \cline{3-11} \cline{4-11} \cline{5-11} \cline{6-11} \cline{7-11} \cline{8-11} \cline{9-11} \cline{10-11} \cline{11-11} 
 & \textbf{DRR (dB)} & 2.1 & \textbf{2.1} & 2.1 & 2.2 & 2.2 & 2.2 & 2.6 & - & -\tabularnewline
\cline{2-11} \cline{3-11} \cline{4-11} \cline{5-11} \cline{6-11} \cline{7-11} \cline{8-11} \cline{9-11} \cline{10-11} \cline{11-11} 
 & \textbf{C5 (dB)} & \textbf{1.9} & \textbf{1.9} & 2.0 & 1.9 & 2.0 & 2.1 & \textbf{1.9} & - & -\tabularnewline
\cline{2-11} \cline{3-11} \cline{4-11} \cline{5-11} \cline{6-11} \cline{7-11} \cline{8-11} \cline{9-11} \cline{10-11} \cline{11-11} 
 & \textbf{Rvol. ($\text{m}^{3}$)} & 4.9 & 4.9 & 5.2 & 4.9 & \textbf{4.8} & 5.4 & 5.1 & - & -\tabularnewline
\cline{2-11} \cline{3-11} \cline{4-11} \cline{5-11} \cline{6-11} \cline{7-11} \cline{8-11} \cline{9-11} \cline{10-11} \cline{11-11} 
 & \textbf{Refc. ($10^{-3}$)} & 56 & 57 & \textbf{52} & 68 & 62 & 65 & 99 & - & -\tabularnewline
\cline{2-11} \cline{3-11} \cline{4-11} \cline{5-11} \cline{6-11} \cline{7-11} \cline{8-11} \cline{9-11} \cline{10-11} \cline{11-11} 
 & \textbf{PESQ} & 0.31 & 0.31 & \textbf{0.30} & 0.30 & 0.34 & 0.38 & 0.39 & 0.66 & -\tabularnewline
\cline{2-11} \cline{3-11} \cline{4-11} \cline{5-11} \cline{6-11} \cline{7-11} \cline{8-11} \cline{9-11} \cline{10-11} \cline{11-11} 
 & \textbf{ESTOI ($10^{-3}$) } & 75 & 76 & 88 & 70 & 68 & \textbf{66} & 74 & 92* & 120\tabularnewline
\cline{2-11} \cline{3-11} \cline{4-11} \cline{5-11} \cline{6-11} \cline{7-11} \cline{8-11} \cline{9-11} \cline{10-11} \cline{11-11} 
 & \textbf{BR (kbs)} & 10.3 & 11.1 & 10.7 & \textbf{10.0} & 86.8 & 85.0 & 10.3 & - & -\tabularnewline
\cline{2-11} \cline{3-11} \cline{4-11} \cline{5-11} \cline{6-11} \cline{7-11} \cline{8-11} \cline{9-11} \cline{10-11} \cline{11-11} 
 & \textbf{SNR (dB)} & 3.5 & 3.6 & 3.7 & \textbf{3.3} & 4.1 & 4.1 & \textbf{3.3} & - & -\tabularnewline
\hline 
\multirow{3}{*}{\textbf{F1 (\%)}} & \textbf{Noise Type } & 66.4 & \textbf{67.5} & - & - & 66.3 & 67.1 & 63.2 & - & -\tabularnewline
\cline{2-11} \cline{3-11} \cline{4-11} \cline{5-11} \cline{6-11} \cline{7-11} \cline{8-11} \cline{9-11} \cline{10-11} \cline{11-11} 
 & \textbf{Codec Type} & 99.7 & 99.5 & 99.6 & \textbf{99.9} & 52.5 & 51.6 & 99.0 & - & -\tabularnewline
\cline{2-11} \cline{3-11} \cline{4-11} \cline{5-11} \cline{6-11} \cline{7-11} \cline{8-11} \cline{9-11} \cline{10-11} \cline{11-11} 
 & \textbf{Overlap Det.} & 92.3 & 92.2 & 92.1 & 92.6 & 94.1 & \textbf{94.2} & 92.1 & - & -\tabularnewline
\hline 
\multicolumn{2}{|c|}{\textbf{\# Param.(M)/RTF($10^{-1}$)}} & 0.97/0.7 & 1.2/0.7 & 0.97/0.7 & 1.2/0.7  & 1.4/12 & 1.6/13 & 0.88/12 & 7.4/2.8 & 1.2/0.4\tabularnewline
\hline 
\end{tabular}
\vspace*{-5pt}
\end{table*}
\begin{table}[H] %[!tbp]
\vspace*{-10pt}
\caption{MAE and confidence intervals~\cite{ConfidenceIntervals} for ACE test set. }
\label{tab:ace-results}
%\centering
\begin{tabular}{|l|c|c|c|}
\cline{2-4} \cline{3-4} \cline{4-4} 
\multicolumn{1}{l|}{} & \textbf{C50 (dB)} & \textbf{T60 (ms)} & \textbf{DRR (dB)}\tabularnewline
\hline 
\multirow{1}{*}{XANE-MFB} & \textbf{2.4(2.2,2.6)} & 176(164,189) & \textbf{1.9(1.8,2.1)}\tabularnewline
\hline 
\multirow{1}{*}{XANE-WLM} & 2.8(2.6,3.0) & \textbf{174(161,187)} & 2.0(1.9,2.2)\textbf{ }\tabularnewline
\hline 
NISA+ & 3.1(2.9,3.3) & 186(172,200) & 3.5(3.3,3.7)\tabularnewline
\hline 
\end{tabular}
\vspace*{-9pt}
\end{table}
\begin{table}[H]
\vspace*{-9pt}
\caption{Clustering results (F1 score, \%) of WLM (original model) and MFB with Conformer based XANE embeddings.}
\label{clustering}
\centering
\begin{tabular}{|c|c|c|c|c|}
\cline{2-5} \cline{3-5} \cline{4-5} \cline{5-5} 
\multicolumn{1}{c|}{} & \textbf{Noise} & \textbf{Reveb.} & \textbf{Overlap} & \textbf{Mean}\tabularnewline
\hline 
WLM & 56.1 & 79.3 & 51.9 & 62.4\tabularnewline
\hline 
XANE & \textbf{90.8} & 96.9 & 98.2 & \textbf{95.2}\tabularnewline
\hline 
XANE-NN & 64.7 & \textbf{100} & 97.4 & 87.4\tabularnewline
\hline 
\end{tabular}
\end{table}
\subsection{Training details}
As shown in Fig.~\ref{fig:nisa_outline}, the output layer in XANE model is divided into separate regression and classification units. We used a Mean Square Error~(MSE) loss for the regression tasks and a cross-entropy loss for the classification tasks. We found that initially training the model on the classification losses only for the first two epochs and then introducing the regression losses helped improve the overall performance of the model, and performed better than the models trained using both losses from the outset. The total loss is, $ L = \lambda_c L_c + \lambda_r L_r$, where $L_c$ and $L_r$ are classification loss and regression losses, respectively. $\lambda_c$ and $\lambda_r$ are the weights for the two losses $L_c$ and $L_r$, respectively. For $\lambda_c$, we use $\lambda_c = 1$, if $0 \leq \mathrm{epoch} < 2$ and $\lambda_c = 0.3$, if $\mathrm{epoch} \geq 2$. Similarly for  $\lambda_r$, we use $\lambda_r = 0$, if $0 \leq \mathrm{epoch} < 2$ and $\lambda_r = 1$, if $\mathrm{epoch} \geq 2$. We placed a threshold of 200~ms (following the value used by the authors of NISA~\cite{nisapp}) on the VAD score as a minimum requirement for estimating other parameters (i.e. if a segment had less than 200~ms speech, the loss function only takes the VAD posterior estimation into account as other parameter estimates would result in a poor quality). We trained the models to 60 epochs using the Adam optimizer with a scheduled learning rate (starting at $10^{-4}$ and 16 epochs patience) with a batch size of 256. Model parameters of the best two checkpoints (measured on the validation set) were averaged to produce the final model. 
\section{Experiments}\label{sec:experiments}
As baselines, we used the recently released model from Torch-Audio-Squim~(TAS)~\cite{kumar2023torchaudio}, which is a deep neural network model trained to estimate PESQ, STOI, and SI-SDR~(objective) and Mean Opinion Score~(MOS), non intrusively. We also implemented the NISA+~\cite{nisapp} system, which uses a mixture of LSTM and Swishnet neural architectures with MelFB and modulation features respectively. We modified the architecture by adding the same 128 dimension embedding and output layer as used by XANE on top of the NISA+ system, to allow training it on all labels and tasks available in our data (we noticed a large degradation in performance when using the original architecture). We additionally used as a baseline,~\textit{STOI-Net}~(SN)~\cite{zezario2020stoinet} which is a deep neural network with a CNN front-end followed by a Bi-directional LSTM~(BLSTM) and self-attention network for STOI estimation (we retrained SN on our training data and modified the output task to predict ESTOI, to allow comparing with NISA+ and our methods). In order to evaluate the quality of the XANE embeddings, we carried out three clustering experiments using the k-means clustering algorithm~\cite{kmeans} and also plotted them using T-SNE~\cite{tsne} dimensionality reduction algorithm, as shown in Fig.~\ref{fig:tsne_conformer}. In the first experiment, we extracted the white noise conditions from the VCTK~\cite{yamagishi2019vctk} based test-set partition (to avoid influence of noise type on the subsequent clustering) and performed a two class clustering using the XANE embeddings, averaged for each utterance. Since, this partition consists of reverberant and coded versions of the data, with and without overlapped speech, we expected this data to partition neatly into the overlapped and non-overlapped classes. 

In the second experiment, we took the union of clean anechoic speech from ACE~\cite{Eaton2016} corpus and the reverberant partition and performed a clustering of the averaged embeddings from each utterance. We expected that the embeddings would cluster into two classes: (a) anechoic and (b) reverberant. Lastly, we experimented with clustering the ACE data according to the noise types in ACE (ambient, babble and fan). The embeddings are made explainable by estimating a large set of acoustic parameters related mainly to the background environment, thus ensuring that the embeddings are modeling information that can be understood in terms of the output metrics. We trained the XANE system with different features and model architectures as described earlier. The input speech was segmented into 1s segments and each segment was processed separately during the training and testing phases. Results on the test set were reported by aggregating the outputs across all the segments. We computed the mean for the regression tasks and used majority voting for the classification tasks. We also performed an ablation study where we evaluated the value of estimating noise type in addition to the other metrics.

\section{Data and Evaluation}\label{sec:data}
We used clean speech from the training partitions of the VCTK~\cite{yamagishi2019vctk} and Timit~\cite{TIMIT} datasets as for synthesising the training data. We convolved the clean speech with RIRs simulated using the image method~\cite{Allen1979}, covering a large configuration of room volumes, reflection coefficients and source-microphone positions. For the overlap conditions, we added speech from a different speaker to the target speaker in an utterance~(3-12~dB range). We then added ambient, babble, white, music and other (mainly domestic noises) noise in 0-30~dB SNR range and processed the audio through the respective CODEC and level augmentation in the range of -0.1 to -10~dBFS. The training dataset was organized into 6 groups, consisting of the three CODEC conditions (uncompressed, Opus ``music'' and Opus ``speech'', 8 to 64~kbps) and the two overlap conditions (overlapped or non-overlapped speech). For each of these groups, we sampled 40k utterances from the clean set and performed the RIR, noise and CODEC augmentations mentioned, resulting in a final set with 240k utterances. For the validation set, we sampled 5\% of this data and removed the corresponding utterances from the training set. Clean speech from the test partitions of the VCTK and Timit datasets was used as base material for synthesising the VCTK test data for the proposed methods and followed the same pattern as the training data, but care was taken to ensure no overlap in speech material or noise and RIR  between the two sets. In addition, we used the ACE~\cite{Eaton2016} dataset for measuring performance of the methods on some of the reverberation metrics as this dataset contains measured RIRs. We use the Mean Absolute Error~(MAE) metric for the regression tasks and F1 score for classification tasks similar to~\cite{nisapp}. We also measured the real-time factor~(RTF) for each of the models on the VCTK-based test data (defined as the ratio of the processing time and the audio length) on a CPU machine with 16~GB RAM.

\section{Results}\label{sec:res}
Figure~\ref{fig:tsne_conformer} illustrates how the XANE embeddings cluster on partitions of the ACE and VCTK test sets for three tasks: reverberant speech detection, overlapped speech detection and noise type classification using the T-SNE algorithm~\cite{tsne}. We can observe that the embeddings for the three tasks cluster very well. We also performed a k-means clustering of the averaged embeddings and measured the F1 scores for the same tasks. We evaluated if the embeddings cluster according to the audio being anechoic or reverberant. As shown in Table~\ref{clustering}, we found that the clustering of this data results in a high separation of utterances belonging to anechoic conditions with the conformer-based model trained without noise type classification XANE-NN) achieving a perfect F1 score. In the second experiment, we evaluated the clustering performance of the embeddings into the three noise types present in the ACE data (ambient, fan and babble), for utterances with SNR less than 20~dB (to exclude clean speech from the analysis). Here, we can see the impact of not including noise type classification in the base model (the XANE-NN system has a poor F1 score of 64.7\%, compared to the full XANE system's 90.8\%). From Fig.~\ref{fig:tsne_conformer} we can observe that the main confusions are between ambient and fan noise, which have similar characteristics. In the third experiment, we took the uncompressed partition of the VCTK based test set and utterances corrupted with white noise (to remove the impact of noise type and CODEC) and separated into a set with overlapping speech and another without overlap. We expected a two class clustering of this data to cluster into two classes, one corresponding to overlapped and the other to non-overlapped speech. The XANE models achieved a high F1 score ($>$97\%) for this task. In contrast, if we clustered based on the WLM embeddings directly, the F1 score was only 51.9. We found that even though the noise type classification performance was poor on the VCTK test set, it did help in the two clustering tasks on ACE data (noise type and reverberant speech classification). In order to make these embeddings explainable, we also extract a number of acoustic parameters and the performance those is presented in Table~\ref{vctk_results} for the VCTK based test set. We found that the MFB features based conformer model performed the best on this test set and removing the noise type estimation also helped improve the performance for a number of regression and classification parameters. Overall, the WLM based models did not perform well on the bit rate and CODEC type estimation tasks, where the conformer model gave an MAE of 85.0~kbps and CODEC type F1 score of just 51.6\%. The task of noise type classification had a poor performance across all methods for this test set. Upon analyzing the confusions for this task it was found that the ambient and other noise types were highly confused and so were the music and babble classes. This points to a shortfall in the labeling of the data and in future work we would aim to address this. Finally, we also evaluated the performance of the C50, T60 and DRR metrics on the ACE corpus, where as shown in Table~\ref{tab:ace-results}, the XANE system outperforms NISA+ by 24.5\% relative MAE and also has a 17x lower RTF.

\section{Conclusions}\label{sec:conclusions}
We presented a novel method for estimating background acoustic embeddings from a speech signal in a non-intrusive manner and making them explainable by estimating a large set of acoustic parameters. Our proposed method shows high accuracy for 11 regression and 3 classification tasks, outperforming the TAS and NISA+ methods, including a 3.4x or 17x lower RTF, respectively. We demonstrate that the embeddings from XANE form well-defined clusters according to the noise type, overlapped speech and reverberant speech detection tasks, with an average F1 score of 95.2\%, compared to 62.4\% observed for the WavLM embeddings. In future work, we will explore the use of XANE embeddings for other speech processing tasks like TTS and improve the noise type labels in the training data.

\bibliographystyle{IEEEtran}
\bibliography{refs}

\end{document}